\documentclass[a4paper]{article}

\usepackage[utf8]{inputenc} 
\usepackage{hyperref}       
\usepackage{url}            
\usepackage{amsfonts}       
\usepackage{amsmath}
\usepackage{graphicx}
\usepackage{caption}
\usepackage{float}
\usepackage[left=3cm,right=3cm,top=2cm,bottom=2cm]{geometry}
\usepackage[english]{babel}
\usepackage{amsmath}
\usepackage{graphicx}
\usepackage[colorinlistoftodos]{todonotes}
\usepackage{float}
\usepackage{dsfont}
\usepackage{bm} 
\usepackage{url}
\usepackage{booktabs} 
\usepackage{multicol}
\usepackage{multirow}
\usepackage{tabularx}
\DeclareCaptionFormat{sanslabel}{#3}%
\usepackage{epstopdf} 
\usepackage{dblfloatfix}

\makeatletter
\def\@seccntformat#1{\@ifundefined{#1@cntformat}%
   {\csname the#1\endcsname\quad}
   {\csname #1@cntformat\endcsname}}
\newcommand\section@cntformat{}     
\makeatother

\begin{document}

\Large
\begin{center}
\vspace{1cm}

\textbf{Optical Incoherence Tomography: a method to generate tomographic retinal cross-sections with non-interferometric imaging systems}


\vspace{0.5cm}

\normalsize
Pedro Mecê$^{1,*}$, Elena Gofas-Salas$^2$, Michel Paques$^{2,3}$, Kate Grieve$^{2,3}$, and Serge Meimon$^4$

\end{center}
\scriptsize
$^1$Institut Langevin, ESPCI Paris, CNRS, PSL University, 1 rue Jussieu, 75005 Paris, France\\
$^2$Quinze-Vingts National Eye Hospital, 28 Rue de Charenton, Paris, 75012, France\\
$^3$Institut de la Vision, Sorbonne Université, INSERM, CNRS, F-75012, Paris, France\\
$^4$DOTA, ONERA, Universit\'e Paris Saclay F-91123 Palaiseau, France\\
$^*$pedro.mece@espci.fr

\normalsize


\begin{abstract} 
Optical tomographic cross-sectional images of biological samples were made possible by interferometric imaging techniques such as Optical Coherence Tomography (OCT) \cite{huang1991optical, mece2020high, mece2020curved}. Owing to its unprecedented view of the sample, OCT has become a gold standard, namely for human retinal imaging in the clinical environment. In this Letter, we present Optical Incoherence Tomography (OIT): a completely digital method extending the possibility to generate tomographic retinal cross-sections to non-interferometric imaging systems such as \textit{en-face} AO-ophthalmoscopes \cite{liang1997supernormal, roorda2002adaptive}. We demonstrate that OIT can be applied to different imaging modalities using back-scattered and multiply-scattered light including systems without inherent optical sectioning. We show that OIT can be further used to guide focus position when the user is ``blind'' focusing, allowing precise imaging of translucent retinal structures \cite{scoles2014vivo}, the vascular plexuses \cite{campbell2017detailed} and the retinal pigment epithelium (RPE) \cite{grieve2018vivo} using respectively split detection, motion contrast, and autofluorescence techniques. 
\end{abstract}

\vspace{1cm}

High-resolution \textit{in-vivo} imaging of the human retina can be achieved using Adaptive Optics (AO) ophthalmoscopes, such as Flood-Illumination Ophthalmoscopes (FIO) \cite{ gofas2018high} and Scanning Laser Ophthalmoscopes (SLO) \cite{roorda2015adaptive}, owing to the capacity of AO to measure and correct for static and dynamic monochromatic ocular aberrations in real-time \cite{jarosz2017high,mece2019higher}. Such high-resolution retinal images play an important role in early-stage retinal disease diagnosis, monitoring the progression of retinal disease and the effect of new therapeutic drugs. \cite{roorda2015adaptive,burns2019adaptive}.

To be able to explore the retinal volume using AO ophthalmoscopes the control of the imaging focal position becomes crucial. At present, the positioning of the image focal plane is typically done empirically by visualizing the \textit{en-face} images displayed in real-time and judging if the retinal structure of interest is sharp or not. This focus guidance approach seems sufficient when using confocal AO-SLO to image hyperreflective retinal layers such as photoreceptors, vasculature, and nerve fiber layer (NFL) \cite{roorda2015adaptive}, especially due to the optical sectioning capability of such imaging systems. Nevertheless, the same cannot be said when using nonconfocal imaging modalities as AO-FIO and AO-SLO split-detection \cite{scoles2014vivo}, multi-offset \cite{rossi2017imaging}, motion contrast \cite{chui2012use}, and autofluorescence \cite{grieve2018vivo}, since displayed images present a weak signal-to-noise ratio (SNR) and a low contrast, and users are mostly ``blind'' focusing. As a result, the acquisition of several image stacks around the retinal layer of interest, \textit{i.e.} for different focal planes, and the assessment of the image quality after the acquisition, to select the best image stack, are mandatory, time-consuming steps which are not always compatible with the clinical environment. To avoid these drawbacks, a focus-guidance tool becomes essential, especially to reveal hypo-reflective or transparent structures such as cone photoreceptor inner segments (IS) \cite{scoles2014vivo}, retinal ganglion cells \cite{rossi2017imaging}, perfusion in microvasculature \cite{gofas2019near, chui2012use} or those masked by neighboring structures of high reflectivity such as RPE lying beneath photoreceptors \cite{grieve2018vivo}. 

Here, we present OIT: a digital method that enables the generation of tomographic retinal cross-sections in non-interferometric AO-ophthalmoscopes. We apply OIT to different AO-ophthalmoscopes modalities: AO-FIO, confocal AO-SLO, split-detection AO-SLO, motion contrast AO-SLO. We demonstrate that most of the retinal layers commonly resolved by OCT cross-sections can also be resolved in OIT cross-sections. Finally, we use OIT to precisely guide focus positioning, enabling imaging of all retinal vascular plexuses, photoreceptor IS and RPE with ease when using respectively motion contrast, split detection, and autofluorescence techniques. 

\begin{figure}[!ht]
\centering
\includegraphics[width=\linewidth]{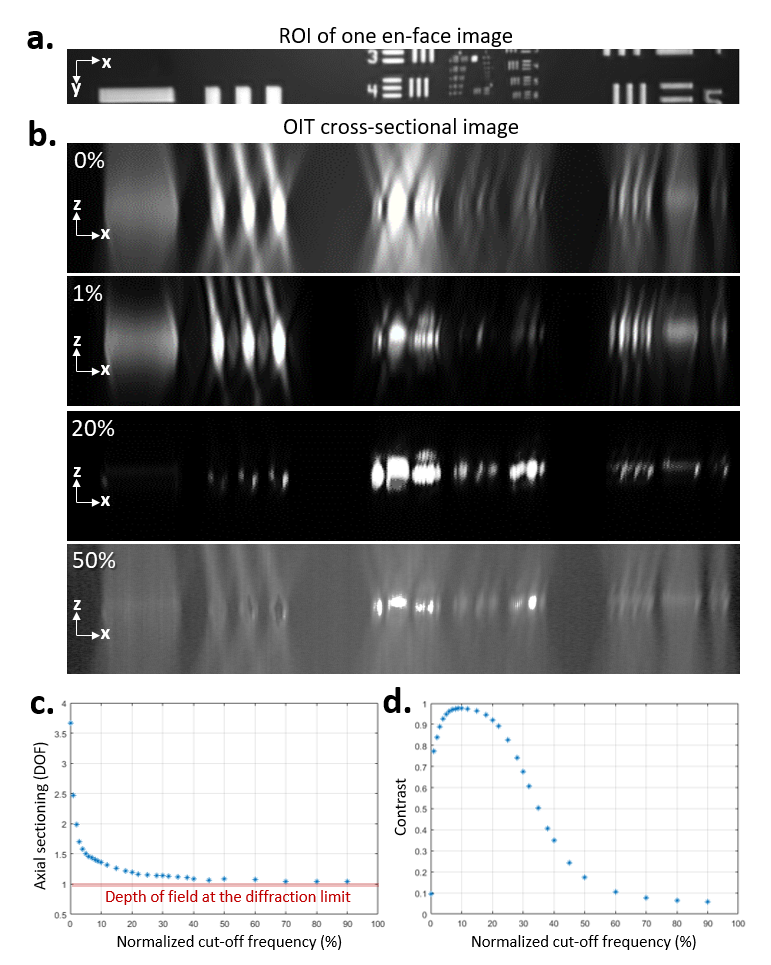}
\caption{\textbf{a.} ROI where the OIT method was applied. \textbf{b.} Generated OIT cross-sections for different normalized cut-off frequencies. \textbf{c,d} Influence of the cut-off frequency choice on, respectively, the axial sectioning capacity (given in terms of the DOF at the diffraction limit) and contrast of OIT cross-section.}
\label{fig:USAF2}
\end{figure}

The OIT procedure is composed of three main steps (see Methods Section for further details): 1) acquisition of \textit{en-face} retinal images from different focal planes, forming a Z-stack; 2) filtering out the low-spatial frequency content of each image using a high-pass filter; 3) Measurement of the image sharpness through the computation of the image energy. A crucial step of the OIT procedure is filtering out the low-spatial frequency content. Indeed, since nonconfocal imaging systems do not present inherent optical sectioning, \textit{i.e.} the capacity to reject out-of-focus photons, the high-pass filter enables one to take advantage of the fact that high-spatial frequency is only present for photons coming from the in-focus plane \cite{lim2008wide}, creating an axial sectioning effect. To demonstrate the axial sectioning capability of the OIT and the impact of the choice of the cut-off frequency, we applied the OIT procedure to a Z-stack obtained from a USAF target using the PARIS AO-FIO. Figures~\ref{fig:USAF2}(a,b) present the region of interest (ROI) of one \textit{en-face} image and OIT cross-sections for different normalized cut-off frequencies, where two behaviors can be noticed. 

On the one hand, by increasing the cut-off frequency, the axial sectioning ability of OIT is enhanced. Figure~\ref{fig:USAF2}(c) outlines this behavior by presenting the axial sectioning as a function of the normalized cut-off frequency. Not surprisingly, the sectioning ability is limited by the depth of field (DOF) at the diffraction limit, as the latter can be defined as the distance from the best focus where the high spatial frequency content starts to lose contrast, and, consequently, the image sharpness decreases \cite{born2013principles}. On the other hand, as the cut-off frequency is increased, the OIT cross-section loses contrast and SNR (Fig.~\ref{fig:USAF2}(d)). This latter effect happens since the contrast is gradually reduced towards zero at a point defined by the lateral resolution of the optical imaging system \cite{born2013principles}. We can expect to obtain sufficient OIT image contrast (higher than 80\%) for normalized cut-off frequencies ranging from 1\% to 25\%. To distinguish different retinal layers, we decided to use a high-pass filter with a normalized cut-off frequency of 20\%, representing a good trade-off between the sectioning ability ($1.2\times$ DOF) and the image contrast (90\%). With this cut-off frequency, for a 7-mm diameter pupil and a light source of 850~nm, we expect to achieve an axial sectioning of $25~\mu m$. As the OIT cross-section is given as a function of the focus position, one can precisely determine beforehand the position of the imaging plane to extract a sharp USAF target image. 

\begin{figure}[!ht]
\centering
\includegraphics[width=\linewidth]{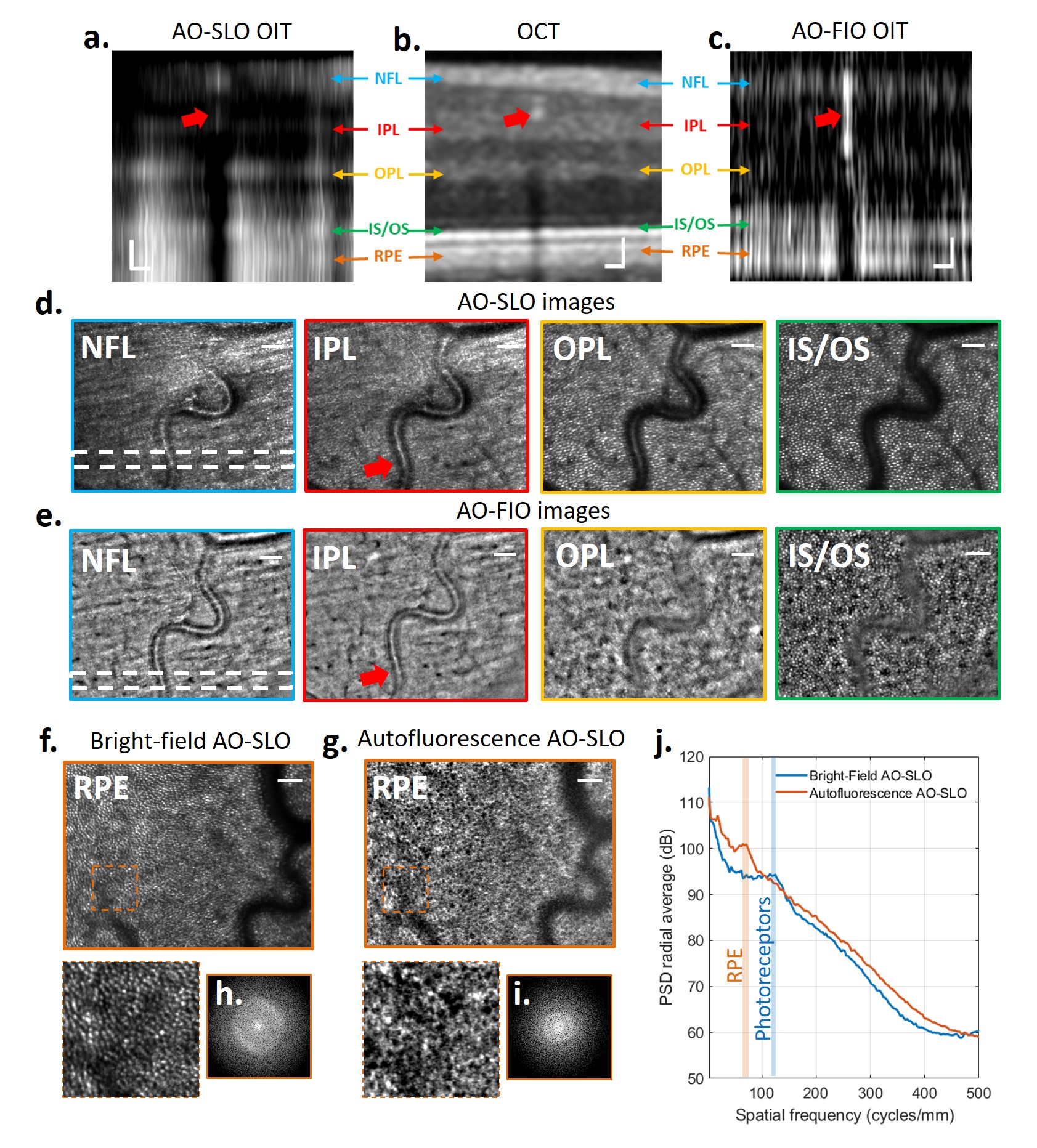}
\caption{\textbf{a-c} Tomographic retinal cross-sections generated by, respectively, AO-SLO OIT, OCT and AO-FIO OIT for the same subject and retinal location, where the main retinal layer can be identified. \textbf{d-g} \textit{En-face} retinal images obtained when precisely positioning the focal plane at the layers labelled in \textbf{a-c}, guided by OIT images. At RPE focal plane, the RPE signal is masked by highly reflective photoreceptors signal due to poor axial resolution, hence need for autofluorescence. \textbf{h,i} Fourier Transform of \textit{en-face} zoomed images (orange dashed-square) \textbf{f,g} at RPE focal plane. \textbf{j.} PSD radial average of \textbf{h,i} outlining the spatial frequency of photoreceptor and RPE mosaic respectively. White-dashed rectangle: ROI where OIT cross-sections were extracted. Red arrows: vessel location. Scale bar: $50 \mu m$.}
\label{fig:FIOvsSLO}
\end{figure}

Figure~\ref{fig:FIOvsSLO}(a,c) presents OIT tomographic retinal cross-sections acquired on a healthy subject at 7$^o$ Nasal using the AO-FIO and AO-SLO, both in bright-field modality. We compare both OIT cross-sections with an OCT image extracted at the same retinal location with Spectralis OCT (Heidelberg Engineering, Germany, Fig.~\ref{fig:FIOvsSLO}(b)). Although OCT can achieve an enhanced axial resolution compared to OIT (\textit{i.e.} not limited by DOF but by the light source bandwidth), most of the retinal layers commonly identified in OCT can be identified in both AO-SLO and AO-FIO OIT cross-sections: 1) the NFL, which gets thicker as eccentricity increases (blue line); 2) two intermediate layers, most probably corresponding to the inner plexiform layer (IPL, red line) and outer plexiform layer (OPL, yellow line); 3) inner/outer segment junction (IS/OS, green line) and RPE (orange line). 

The proposed retinal layer labeling can be further confirmed by looking at the \textit{en-face} images acquired when positioning the imaging focal plane at each of these layers (Figs.~\ref{fig:FIOvsSLO}(d,e)). The RPE labeling is confirmed when applying autofluorescence with AO-SLO \cite{grieve2018vivo} at the given focus position (Figs.~\ref{fig:FIOvsSLO}(f-i)). Figure~\ref{fig:FIOvsSLO}(j) presents the radial averaged power spectral density (PSD) for both bright-field and autofluorescence AO-SLO acquired simultaneously, presenting the typical spatial frequency of, respectively, the photoreceptor and RPE mosaics. We measured a photoreceptor density of $18~000~cells/mm^2$ and an RPE density of $5~000~cells/mm^2$, which is consistent with previous studies for the given retinal eccentricity \cite{grieve2018vivo,cooper2016evaluating}. The ROI where the OIT tomographic retinal cross-sections were generated, comprising a retinal vessel, is indicated by the white dashed-rectangle in NFL \textit{en-face} images in Figs.~\ref{fig:FIOvsSLO}(d,e). Through the OIT cross-sections, it is possible to identify the vessel present in the ROI and the OCT (red arrows) and conclude on its axial position, here at the superficial vascular plexus (SVP) \cite{lavia2020retinal}. One main advantage of the OIT compared to OCT is its enhanced lateral resolution, enabling visualization of some retinal features that cannot be visualized in OCT, such as interconnecting capillaries, linking different capillary plexus \cite{campbell2017detailed} (Supplementary~Video~1).  

\begin{figure}[!ht]
\centering
\includegraphics[width=\linewidth]{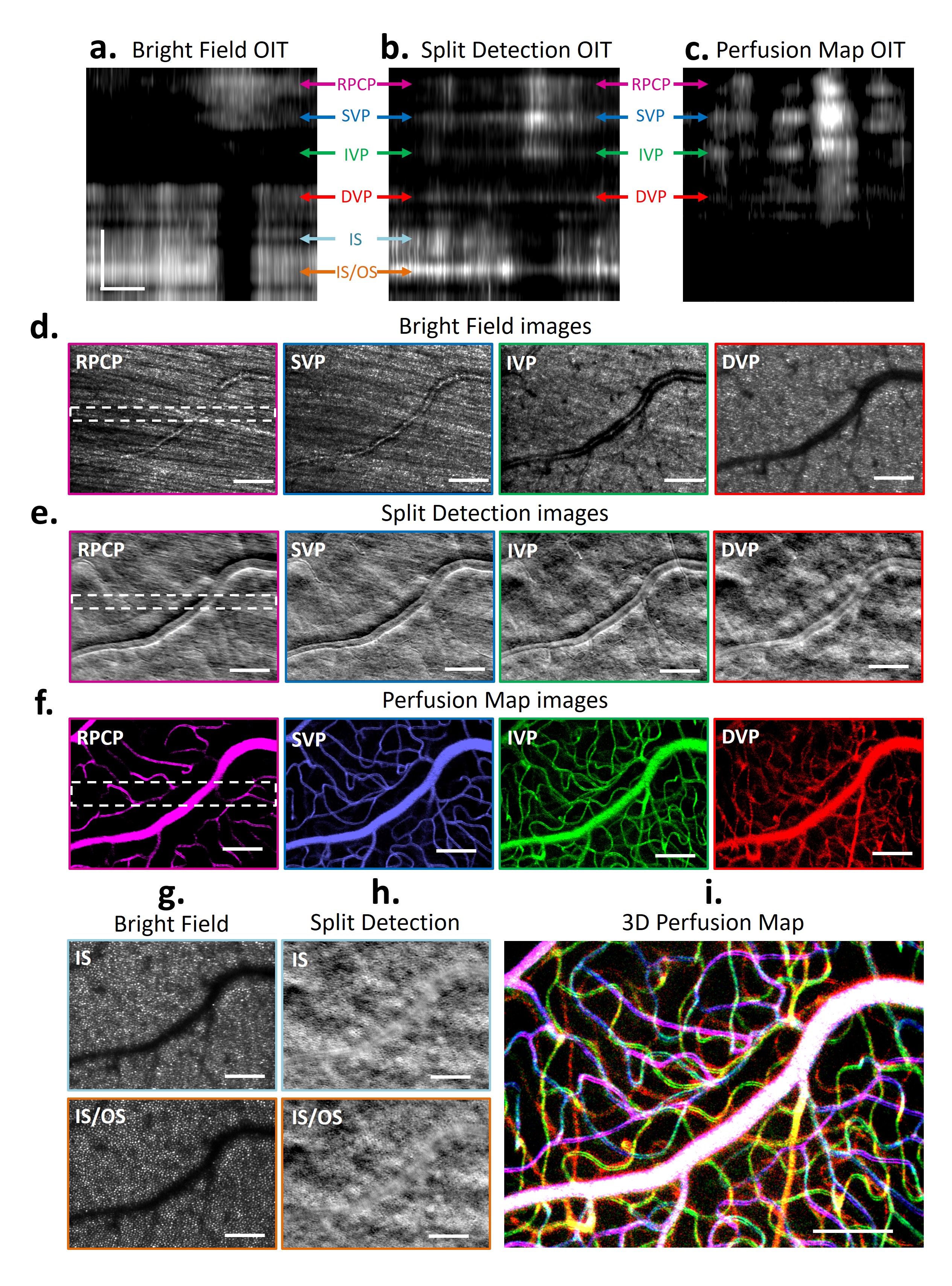}
\caption{\textbf{a-c} Tomographic retinal cross-sections generated by, respectively, bright-field, split detection and motion contrast techniques in AO-SLO for the same subject at 7$^o$ Nasal where the NFL is dense and four vascular plexuses can be seen. \textbf{d-h} \textit{En-face} retinal images obtained when precisely positioning the focal plane at the layers labelled in \textbf{a-c}, with the help of OIT method. \textbf{i} Composite perfusion map image, revealing the 3D organization of the retinal vascular network. White-dashed rectangle: ROI where OIT cross-sections were extracted. Scale bar: $100 \mu m$.}
\label{fig:Nasal}
\end{figure}

Multiply scattered light (or nonconfocal) imaging modalities such as dark-field, offset aperture and split detection, largely applied in AO-SLO \cite{burns2019adaptive}, and recently introduced for AO-FIO \cite{meimon2018manipulation,gofas2019near}, provide excellent contrast for blood vessels and mural cells \cite{chui2012use,gofas2019near} and translucent retinal structures \cite{scoles2014vivo,rossi2017imaging}, all poorly or not visualized in back-scattered light imaging systems. Because the coherent detection of OCT limits the use of multiply scattered light, OCT is not able to generate split detection tomographic retinal cross-sections. On the other hand, the OIT procedure can be applied to Z-stacks acquired in multiply scattered light imaging modalities to generate split detection tomographic retinal cross-sections, revealing a different cross-sectional view of the retina. Figures~\ref{fig:Nasal}(a,b) present a comparison between OIT cross-sections generated through bright-field confocal AO-SLO and nonconfocal split detection AO-SLO at 7$^o$ Nasal of a healthy subject. Colored arrows in OIT cross-sections indicate the focus position where en-face images, presented in Fig.~\ref{fig:Nasal}(d,e), were acquired. White dashed rectangles indicate the ROI where OIT cross-sections were generated. Full Z-stacks can be visualized in Supplementary~Video~2.  

\begin{figure}[!ht]
\centering
\includegraphics[width=\linewidth]{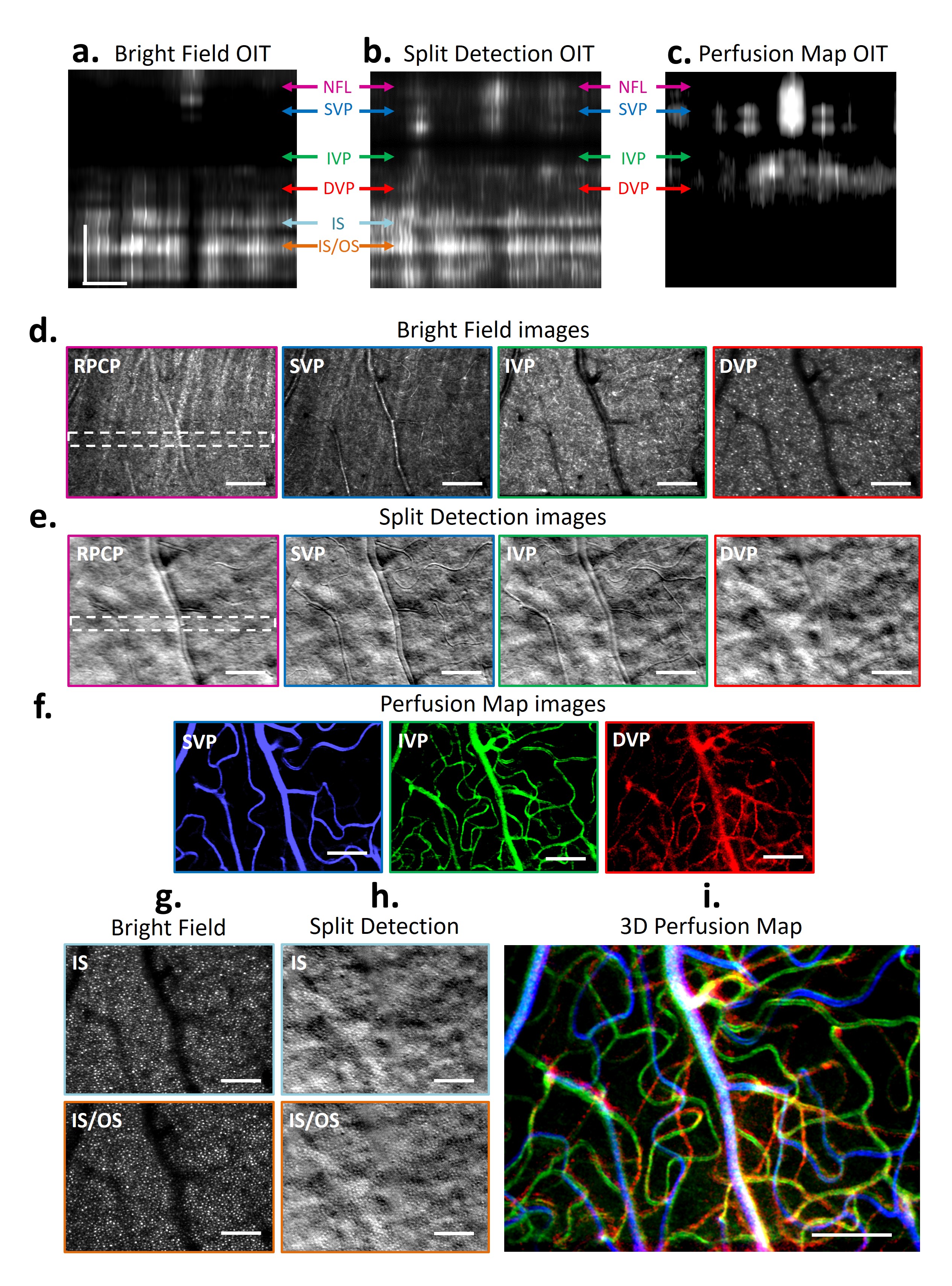}
\caption{\textbf{a-c} Tomographic retinal cross-sections generated by, respectively, bright-field, split detection and motion contrast techniques in AO-SLO for the same subject at 7$^o$ Temporal where NFL is less thick and three vascular plexuses can be seen. \textbf{d-h} \textit{En-face} retinal images obtained when precisely positioning the focal plane at the layers labelled in \textbf{a-c}, with the help of OIT method. \textbf{i} Composite perfusion map image, revealing the 3D organization of the retinal vascular network. White-dashed rectangle: ROI where OIT cross-sections were extracted. Scale bar: $100 \mu m$.}
\label{fig:Temporal}
\end{figure}

Three main differences can be noticed when producing split detection OIT cross-sections compared to those generated from bright-field images. The first difference concerns the NFL layer which seems to disappear in split detection OIT, indicating, as previously stated, that NFL becomes mostly transparent in multiply scattered light modalities \cite{chui2012use}. Secondly, retinal layers in the inner retina get brighter with split detection OIT. By producing the OIT corresponding to a Z-stack of perfusion map images (Fig.~\ref{fig:Nasal}(c)), we can deduce that these layers correspond to vascular plexuses, of which we expect there to be four at 7$^o$ Nasal \cite{campbell2017detailed,lavia2020retinal}. The focus position was adjusted using split detection OIT to acquire perfusion map images of each of the four vascular plexuses (Fig.~\ref{fig:Nasal}(f)), named according to \cite{campbell2017detailed}: radial peripapillary capillary plexus (RPCP), SVP, intermediate vascular plexus (IVP) and deep vascular plexus (DVP). Owing to the precise location of focus position, one can generate depth-color coded perfusion maps with ease, revealing the 3D organization of the vascular network (Fig.~\ref{fig:Nasal}(i)). Finally, another interesting finding is a retinal layer, just above the IS/OS, that gets brighter with split detection OIT. By using OIT to position the focal plane at this layer we were able to precisely image photoreceptor IS (Figs.~\ref{fig:Nasal}(g,h)) \cite{scoles2014vivo}. Figure~\ref{fig:Temporal} and Supplementary~Video~3 present the same comparison and results but at 7$^o$ Temporal, where NFL is less dense and all three expected vascular plexuses \cite{lavia2020retinal} and the photoreceptor IS layers are visible in split-detection OIT, enabling focus-guidance and image acquisition of perfusion maps (for vascular plexuses) and the IS. 

Throughout this study, we showed the capacity of OIT to generate tomographic retinal cross-section for various non-interferometric imaging modalities. While this Letter discusses the application of OIT to retinal imaging, this method can be applied to other samples (\textit{e.g.} cornea \cite{jalbert2003vivo} and skin \cite{rajadhyaksha2017reflectance}) and for other high-resolution microscopic/imaging techniques that go beyond biological samples. Since the focus position is known, OIT can be used as a focus guidance tool to image a retinal layer of interest, even when the user is ``blind'' focusing. We demonstrate this asset by extracting perfusion maps from all vascular plexuses, photoreceptor IS and RPE images assisted by OIT using respectively motion contrast, split detection, and autofluorescence techniques. Moreover, the split detection cross-sectional view of the retina, not possible with OCT, may be a valuable tool to understand the origin of retinal features observed in multiply scattered light modalities \cite{guevara2020origin}. Since OIT is a completely digital method, it can be easily implemented in any optical imaging system. Moreover, an OIT cross-section can be obtained in a relatively short time, even for AO-SLO, as the ROI of \textit{en-face} images from the Z-stack is reduced in one direction (about $50~\mu m$), which can coincide with the slow galvanometer scanner axis. Thanks to its easy implementation, OIT can be used during imaging sessions to help focus guidance. One can start by acquiring a fly-through focus movie, generate the patient OIT image, then using the cross-sectional view and focus information to precisely position the imaging focal plane at the retinal layer of interest, avoiding loss of time in acquiring images from different depths and selecting the best image sequences afterward. Finally, laser photocoagulation retinal surgery can also benefit from OIT, by precisely focusing the therapeutic laser at the diseased tissue avoiding damaging neighboring healthy tissues \cite{mece2018can}. 

\section*{Acknowledgements}

This work was supported by Agence Nationale de la Recherche CLOVIS3D grant (ANR-14-CE17-0011), and European Research Council HELMHOLTZ grant (\#610110). The authors want to thank Laurent Mugnier, Cyril Petit, Yann Lai-Tim and Antoine Chen for fruitful discussions.

\section*{Author contributions}

P.M. wrote the software for data processing and generation of OIT cross-sections, analyzed the results and drafted the manuscript. P.M., E.G. and K.G. designed the experiment. P.M. and S.M. developed the presented method. P.M., E.G., K.G. and M.P. collected data. M.P., K.G. and S.M. provided overall guidance to the project. All authors discussed the results, reviewed and edited the manuscript.

\section*{Competing interests}
P.M. and S.M. are listed as inventors on a patent application (FR 1904271) related to the work presented in this manuscript. All other authors have nothing to disclose.

\newpage

\newpage

\newpage

\newpage

\newpage
\section*{Methods}

\subsection*{Optical Incoherence Tomography procedure}
As mentioned in the Letter, the OIT procedure is composed of three steps: Z-stack acquisition, image filtering and computation of the image energy. 
\paragraph{Z-stack acquisition:} The acquisition of \textit{en-face} images from different focal planes, forming a Z-stack (or a fly-through movie), can be done in two different ways: step-wise, by manually changing the focus position and acquiring enough images for each imaging plane; or continuously, by operating a fly-through focus (continuous change of focus position) during imaging acquisition. While the OIT cross-section generated by the former will present a better SNR and contrast, as multiple images for each focus position can be averaged, the latter will be faster. Although in this Letter we only used the step-wise acquisition, the fly through focus acquisition might be sufficient depending on the retinal structure of interest or when the SNR is high, as in the case, for example, of confocal AO-SLO.

\paragraph{Image filtering:} After acquiring the Z-stack, each image has its low-spatial frequency content filtered out. Here, we empirically chose to use an order 2 Butterworth filter to avoid adding high-spatial frequency artifacts to images. 

\paragraph{Image energy computation:}
To obtain OIT cross-sectional images, similar to an OCT "B-scan", each image is divided into an overlapping grid of $n \times m$ pixel ROI, where each ROI is displaced from the previous by 1 pixel. To favor a trade-off between point-wise accuracy and smoothness, we empirically chose $m = 4$ and $n = 80$, which is equivalent of $3 \mu m \times 60 \mu m$. The image energy of each ROI is then computed as follows:

\begin{equation}
    \sum_{i=1}^m\sum_{j=1}^n|\widetilde{I}(i,j)|^2
\end{equation}
where $\widetilde{I}$ is the Fourier Transform of the filtered ROI of $m \times n$ dimensions. High energy values will be obtained in ROI presenting a significant amount of high-spatial frequency content. The axial sectioning capacity of OIT is limited by the DOF at the diffraction limit which can be defined, according to \cite{born2013principles} as: 
\begin{equation}
    DOF = \frac{n\lambda}{NA^2}
\end{equation}
Finally, to facilitate visualization of OIT cross-sections, we used bicubic interpolation in the axial direction.

\subsection*{Experimental set-up}
The Z-stacks necessary to generate OIT cross-sections were obtained using the PARIS AO-FIO and a modified version of the MAORI (multimodal adaptive optics retinal imager) AO-SLO (Physical Sciences, Inc., Andover, MA, USA). Both systems were described in detail elsewhere \cite{gofas2018high, grieve2018vivo}. The Z-stack from the PARIS AO-FIO was obtained by translating the imaging camera parallel to the optical axis with a constant step of 30~$\mu m$ in the retinal plane. AO-SLO Z-stacks were obtained by adding constant defocus values to the deformable mirror (equivalent to an axial displacement of 20~$\mu m$ of the retinal plane).

\subsection*{Subjects}
Image acquisition was performed on two healthy subjects aged 25 and 38. Research procedures followed the tenets of the Declaration of Helsinki. Informed consent was obtained from subjects after the nature and possible outcomes of the study were explained. The study was authorized by the appropriate ethics review boards (CPP and ANSM (IDRCB numbers: 2016-A00704-47 and 2019-A00942-55)). Before the acquisition, pupil dilation and accommodation paralysis were performed by introducing one drop of each Tropicamide and Phenylephrine 10\%, assuring a constant pupil diameter and minimal interference of defocus dynamics due to natural accommodation during Z-stack acquisition \cite{mece2019higher,mece2019visualizing}. Subjects were seated in front of the system and stabilized with a chin and forehead rest and asked to fixate a target placed at an infinite focal conjugate.

\subsection*{Imaging acquisition}
For each focal plane, a total of 100 images were acquired. When using the AO-SLO device, four sets of data were recorded simultaneously:  bright-field, split-detection, and autofluorescence imaging; and the deformable mirror focal plane position. In this configuration, the AO-SLO light level was 1.65 mW. For the PARIS AO-FIO, the total power entering the eye from the illumination source and the wavefront sensor laser beacon were respectively 350~$\mu W$ and 1.8~$\mu W$. For both imaging systems, the light level was below the ocular safety limits established by the ISO standards for group 1 devices. 

\subsection*{Imaging processing}
Following the acquisition, to correct for fixational eye movements \cite{mece2018fixational}, strip-based registration (only for the AO-SLO images) and normalized cross-correlation registration (for both devices) were performed on each image sequence for a given focal plane. After registration, we selected 20 out of 100 images presenting the best quality (computed through the image energy \cite{mece2019higher}). Then the 20 selected images were averaged or had the temporal standard deviation computed to extract perfusion maps, providing the final Z-stack used to generate the OIT retinal cross-sections. Before performing the OIT procedure, images composing the Z-stacks were registered pairwise. 

\subsection*{Data availability}
The study data are available from the corresponding author upon request.

\subsection*{Code availability}
The software to generate OIT cross-section is available upon request.
\newpage
\bibliographystyle{ieeetr}
\bibliography{OIT}

\newpage
\section{Supplementary information}
\paragraph{Video 1} Interconnecting capillaries are visible in OIT cross-section generated with AO-FIO. Upper image: ROI used to compute the OIT cross-section. Lower image: Corresponding OIT cross-section. Arrows highlight the three-dimensional position of interconnecting capillaries. Blue arrow: capillaries connecting the superficial vascular plexus (SVP) and the deep vascular plexus (DVP). Red arrow: capillaries connecting the SVP and the intermediate vascular plexus (IVP). Green arrow: capillaries connecting the SVP with the radial peripapillary capillary plexus (RPCP). Retinal image acquired at 7$^o$ Nasal.

\paragraph{Video 2} The fly-through focus movies for bright-field AO-SLO and split-detection AO-SLO modalities at 7$^o$ Nasal.

\paragraph{Video 3} The fly-through focus movies for bright-field AO-SLO and split-detection AO-SLO modalities at 7$^o$ Temporal.

\end{document}